\documentclass[%
 ,twocolumn%
 ,secnumarabic%
,superscriptaddress, footinbib,
,amssymb, amsmath,nobibnotes, aps, prl]{revtex4-1}
\usepackage{bm}%
\usepackage{graphicx}
\usepackage{epstopdf}
\begin{document}

\title{Elastic building blocks for confined sheets}%

\author{Robert D. Schroll}
\affiliation{Physics Department, University of Massachusetts, Amherst MA 01003}
\author{Eleni Katifori}
\affiliation{Center for Studies in Physics and Biology, Rockefeller University, New York, NY 10065}
\author{Benny Davidovitch}
\affiliation{Physics Department, University of Massachusetts, Amherst MA 01003}
\date{\today}%

\begin{abstract}
We study the behavior of thin elastic sheets that are bent and strained under the influence of weak, smooth confinement.
We show that the emerging shapes exhibit the coexistence of two types of domains that differ in their characteristic stress distributions and energies, and reflect different constraints. A focused-stress patch is subject to a geometric, piecewise-inextensibility constraint, whereas a diffuse-stress region is characterized by a mechanical constraint - the dominance of a single component of the stress tensor.
We discuss the implications of our findings for the analysis of elastic sheets that are subject to various types of forcing.
\end{abstract}

\maketitle

The geometry and mechanics of elastic sheets have recently become a focus of intense activity for chemists, biologists, engineers, and physicists \cite{Genzer06,Witten07}. This interest has been driven in part by studies that demonstrated its relevance to biological tissues \cite{Sharon04, BenAmar08}, 
and by technological advance that enabled the production of extremely thin films with precise material properties \cite{Bowden98,Huang07}. Such thin sheets undergo a buckling instability even under minor compressions, and their typical state is therefore far above threshold, where the energetic cost of straining is much larger than that of bending \cite{Witten07}. In such situations traditional perturbation methods that are used to analyze sheets close to buckling threshold \cite{LL86} are no longer available, and a full nonlinear analysis of the system
is required. A major theoretical challenge here
is the development of a formalism that effectively addresses the configurations realized by elastic sheets as their thickness becomes exceedingly small. A central question for such a theory is what are the basic ``building blocks" that compose these asymptotic shapes.

Singular types of building blocks include developable cones (``d-cones'') \cite{BenAmar97,Cerda99} and ``minimal ridges'' \cite{Lobkovsky96, Venkataramani04}, which are reflected in the branched network of vertices and sharp folds in a crumpled paper \cite{Witten07}.
In the limit of an infinitely thin sheet, these asymptote to points 
or lines. 
This behavior manifests a consequence of Gauss's \emph{Theorem Egregium},
according to which patches that are curved in two directions must be strained \cite{Witten07}.
These singular structures focus elastic energy in small regions that are highly bent 
 and hence strained, creating an asymptotically {\emph{piecewise-inextensible}} (origami-like) shape in which the rest of the sheet remains unstrained in flat facets.
 Stress focusing has been a subject of many 
 studies in the last two decades \cite{Witten07}. However, it has become clear that shapes of thin sheets are not fully describable by the stress focusing idea. 
 For example, it is known that uniaxial tension leads to smooth wrinkling patterns that are curved (hence strained) everywhere in both directions \cite{Cerda03,Huang10}. Even without exerted tension, certain boundary conditions \cite{MahaEPL} lead to shapes that reflect a smooth distribution of strain and curvature. It was even proposed that stress focusing may appear only under large confinement \cite{AudolyBook} or in response to sharp boundaries \cite{Venkataramani04}, although recent results may suggest that this is not the case \cite{Aharoni10}.
 We are thus led to ask a number of fundamental questions: What type of boundary conditions yield singular structures?
 Are there other fundamental structures that are necessary for describing the configurations of thin sheets?

 Motivated by these questions, we study in this Letter a sheet under simple confinement, representative of the general class of boundary conditions that are not ``tailored" to yield a piecewise-inextensible shape.
 Our results render three important messages: First, we show that a focused-stress structure appears even under weak, smooth confinement. Second, a focused-stress zone may coexist with a large region in which energy is smoothly distributed.
 This ``\emph{diffuse-stress}'' zone constitutes a new, so far largely overlooked, building block.
 Third, in contrast to stress focusing, diffuse-stress domains are not dominated by a geometric 
 constraint but rather by a mechanical one: vanishing ratio between compressive and tensile components of the stress tensor.
Our observations suggest a cornerstone for a theory of the asymptotic shapes of thin sheets under general conditions.

Our system, Fig. 1a, consists of a semi-infinite rectangular sheet of thickness $t$ where one long edge, say $y\!=\!W$, is displaced inward by $\tilde{\Delta}W$ with $\tilde{\Delta} \!\ll \!1$.
\begin{figure}
\includegraphics{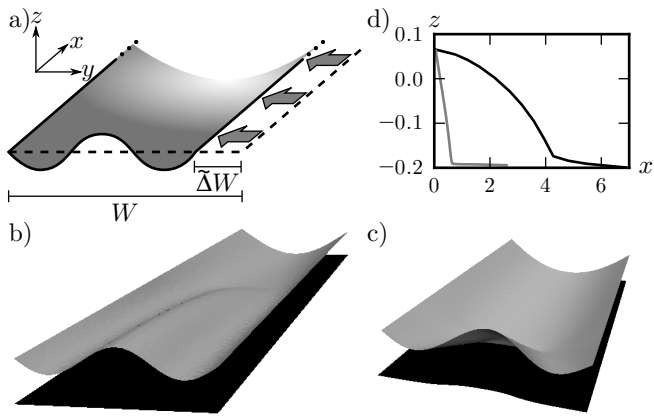}
\caption{(a)~A semi-infinite sheet of width $W$ confined in one direction by a distance $\tilde\Delta W$.  The near end of the sheet is forced into a 3-buckle shape in the $y$-$z$ plane.
(b)~The shape of a sheet with $t/W = 0.002$, $\tilde \Delta = 0.1$, and $\nu=\frac{1}{3}$. Note the extended smooth area near the prescribed edge, which eventually terminates in a focused structure. (c)~When the prescribed edge is allowed to be non-planar, the sheet has only a focused stress region.  (d)~The centerline profiles for the shapes in (b) (black) and (c) (gray).}
\end{figure}
Far from the edge at $x\!=\!0$, a sufficiently thin sheet ($t\ll \sqrt{\tilde{\Delta}} W$) would naturally buckle to an asymptotically $x$-independent shape $\zeta_1(x,y) \! = A_1 \cos(\pi y/W)$. At $x\!=\!0$ we impose a ``3-buckle" profile $\zeta_3(x\!=\!0,y) \!= A_3 \cos(3\pi y/W)$, required to lie in the $y$-$z$ plane. Absence of strain at the boundaries determines the amplitudes: $A_j \approx \! 2 \sqrt{\tilde{\Delta}} W/ j \pi \ (j=1,3)$.
The transition between the 3-buckle and 1-buckle shapes requires the formation of a strained region, where curvature exists in both directions. It is the structure of this strained region that we address here. Our choice of boundary conditions is motivated by several reasons:
(\emph{a}) The strainless 1-buckle and 3-buckle shapes are, respectively, the asymptotic ground state and a low-energy meta-stable state of the bending energy (\ref{energy}) under one-dimensional confinement. Similarly to the universal nature of phase transformation, e.g.~between solid and super-cooled liquid, the transition between the strainless states can be expected to be a generic form for accommodating unavoidable strain, rather than a shape dependent on the specific boundary conditions. In contrast, other studies directly induce strain by pinching \cite{MahaEPL,MahaPRL} or flattening \cite{Sternberg01,BenBelgacem00} an edge of a confined sheet.
(\emph{b}) Despite the nontrivial structure of the strained region,
the simplicity of our system allows quantitative description both numerically and analytically. This enables us to clearly distinguish between ``diffuse-stress" and ``focused-stress" types of building blocks, and to formulate general asymptotic conditions that could be used under more complicated constraints.
(\emph{c}) Beyond the general lessons drawn from this system, it deserves its own right as a natural ``unit cell" of hierarchical, multi-scale patterns on elastic sheets 
\cite{Sternberg01,BenBelgacem00,Conti05,Huang10,Davidovitch09}.

The configuration of the sheet is found by minimizing the F\"oppl\textendash von K\'arm\'an (FvK) elastic energy, $U = U_S + U_B$, which contains stretching and bending terms \cite{LL86}:
\begin{align}
U_S &= \frac{1}{2}  \int dx  dy  \  \sigma_{ij} u_{ij} \ ;  \ U_B = \frac{1}{2} B   \int dx  dy \  (\nabla^2\zeta)^2 \label{energy} \\
\sigma_{ij} &= \frac{Y}{1-\nu^2}[(1\!-\!\nu)u_{ij}+\nu\delta_{ij}u_{kk}] .
\label{stressstrain}
\end{align}
$Y \!=\! Et$ and $B \!=\! Et^3/12 (1-\nu^2)$ are, respectively, the stretching and bending modulii, $E$ is the Young modulus, and $\nu$ is the Poisson ratio.
The (geometric) nonlinearity of $U$ is associated with the strain $u_{ij} = \frac{1}{2}\ (\partial_iu_j +\partial_ju_i + \partial_i\zeta\partial_j\zeta)$.
We use the numerical software Surface Evolver \cite{Brakke92} to minimize this energy on a rectangular sheet.
One end of the sheet is fixed to the 3-buckle shape while the other is free.  The sheet is long enough that the free end takes the shape of a single buckle.  No bending moment is applied to the long edges.

\begin{figure}
\includegraphics{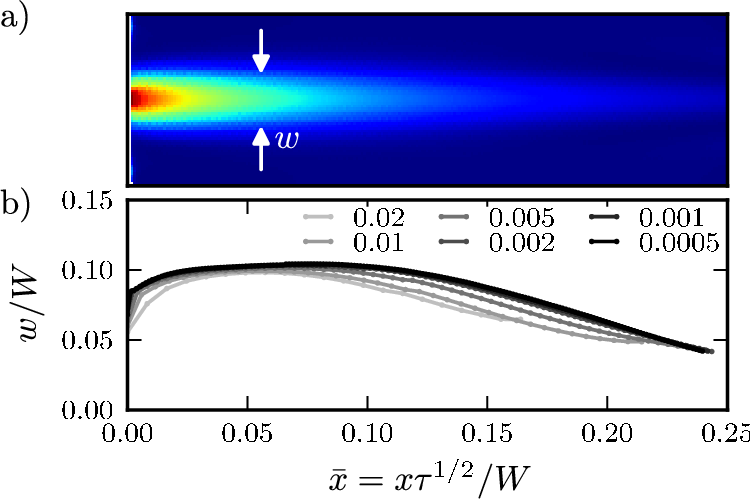}
\caption{(a) The density of stretching energy in the diffuse-stress region of Fig.~1b.
The prescribed edge is at the left.  The stretching energy density can be well approximated by the form $A(x)\exp[-(y/w(x))^2]$ .
(b) When plotted against $\bar x=x \tau^{1/2}/W$, the width $w(\bar x)$ does not depend on thickness ($t/W$ as indicated).
}
\end{figure}
A representative shape, shown in Fig.~1b, exhibits two prominent features.
First, the transition terminates sharply at a small, stress-focusing zone, beyond which the single buckle shape is approached.
Notably, this focused-stressed structure appears under weak, smooth confinement, unlike the d-cones and ridges of \cite{Lobkovsky96,Cerda99,Venkataramani04}.
Second, the transition between the two states occurs over a large distance $L_t$, which
diverges as $t^{-1/2}$ as $t$ is reduced. 
Similar scaling was observed in a pinched cylindrical shape \cite{MahaEPL,dePablo03}, and was shown to arise from a competition between two dominant energies: excess bending (favoring small $L_t$) and stretching (favoring large $L_t$). A similar type of energetic balance has been noted already in \cite{Sternberg01,BenBelgacem00,Conti05}.
The bending energy is dominated by the curvature in the $\hat y$ direction, $\kappa \sim \sqrt{\tilde{\Delta}}/W$, while the in-plane stretching is dominated by strain along $\hat x$, estimated as $u_{xx} \sim (\frac{\partial\zeta}{\partial
x})^2 \sim \tilde{\Delta}(W/L_t)^2$.
Balancing the two energies $B \kappa^2 \sim Y u_{xx}^2$, we obtain
\begin{equation}
L_t \sim W /\tau^{1/2} \  \  \text{and} \  \ U \sim E W^3 (\tau \tilde{\Delta})^{5/2} \ ,
\label{diffusedstressLtenergy}
\end{equation}
where $\tau \equiv t/\left(W \sqrt{\tilde{\Delta}}\right)$ is an effective (dimensionless) measure of the thickness. Notice that far from buckling threshold $\tau \!\ll\!1$.
Notably, this argument suggests that in the transition region stress is not focused; instead stress is smoothly distributed over area $W L_t$ that diverges as $t\!\to\! 0$.
Figure 2 demonstrates that the lateral extent $w(x)$ of the highly stressed region is a finite, thickness independent fraction of the width $W$, while the length scales with $L_t$, as seen in the rescaling with $\bar x \equiv x\tau^{1/2}/W$.
We therefore identify this as a diffuse-stress region.
\begin{figure}
\includegraphics{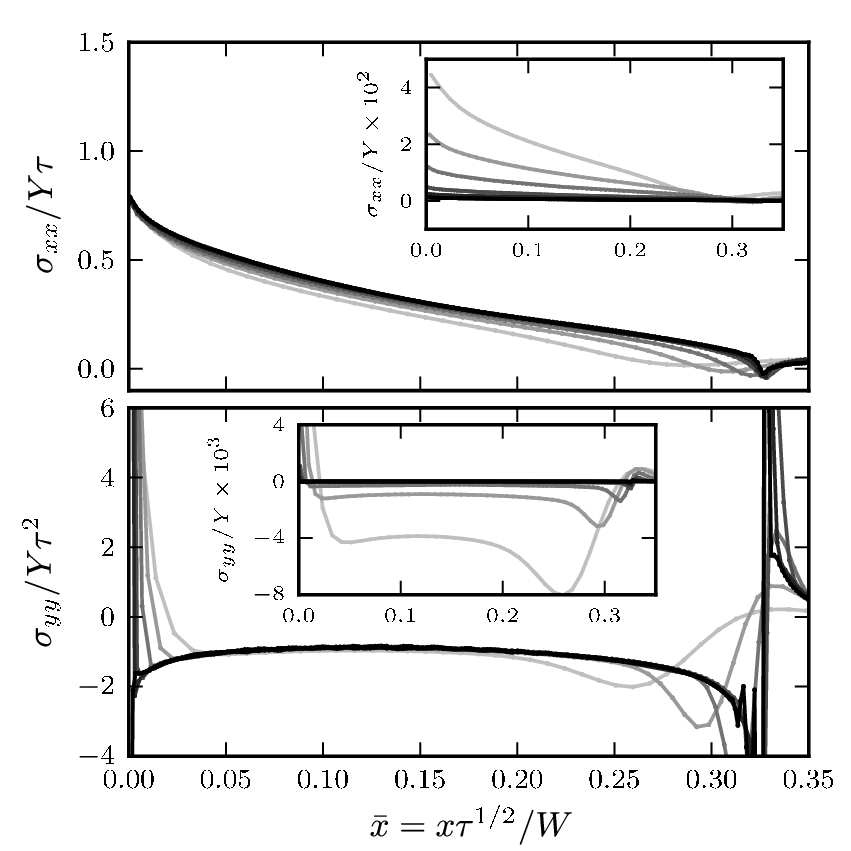}
\caption{The diagonal stress components $\sigma_{xx}$ and $\sigma_{yy}$, measured along the centerline of the shapes and plotted against the scaled $\bar x$, for the simulations of Fig.~2b (same legend).  The data collapse when $\sigma_{xx}$ is rescaled by $\tau$ (top) and $\sigma_{yy}$ by $\tau^2$ (bottom).  Inset, the unscaled stresses, also plotted against $\bar x$.}
\end{figure}

Naively, this observation seems to be inconsistent with the previously-noted stress focusing apparent near $x \!\approx\! L_t$. However, careful examination of this area resolves this paradox: The diffuse-stress region terminates at $\bar{x}^{*} \approx 0.32$, beyond which a typical focused-stress structure appears.
This structure resembles a d-cone whose characteristic features become sharper as $t$ is reduced, consistently with known scaling laws \cite{Cerda99,MahaPRL}.
In contrast to the diverging size $WL_t$ of the diffuse-stressed region, the focused-stress structure is confined to an area $\sim W^2$.

 Our results demonstrate the emergence of a focused-stress domain in response to smooth strainless boundary conditions, and its coexistence with a large diffuse-stress region.
Evidence for the general occurrence of such shapes can be obtained by allowing the planar edge profile to rotate an arbitrary angle $\theta$ around the $y$ axis [$u_x(x\!=\!0,y) = A_3 \cos\theta \cos(3\pi y/W)$].
 While this freedom does slightly modify the shape, we found no qualitative change of the above picture unless the edge was allowed to assume a nonplanar shape [i.e. an arbitrary $u_x(x\!=\!0,y)$].
In this case, shown in Fig.~1c, only a focused-stress structure appears near the edge, giving way to the single buckle after a distance $L_t\sim W$. This length does not depend on thickness, reflecting the geometrical nature of the inextensibility constraint \cite{dconenote}.
The energy associated with this focused-stress structure is observed to scale with $\tau^{8/3}$ \cite{inprep}, similar to the energy scaling of minimal ridges \cite{Lobkovsky96,Witten07}.  This energy is asymptotically negligible relative to the diffuse-stress energy, (\ref{diffusedstressLtenergy}).
 The ``fine-tuning'' required to eliminate the diffuse-stress region suggests that although stress focusing is energetically favorable, it is generally insufficient to relieve the strain in a confined sheet. The formation of a diffuse-stress region seems to be the ``second best" alternative for an unavoidable stretching, and hence we conjecture that a coexistence of these two building blocks, focused-stress and diffuse-stress, is a general feature of very thin sheets under arbitrary confinements.
 

The markedly different nature of diffused-stress and focused-stress domains and their coexistence in a single shape lead one to expect that an analytic computation of the shapes of confined sheets requires the matching of two distinct types of asymptotic expansions. 
The asymptotic nature of focused-stress structures is known to reflect a geometric principle \cite{Witten07}: a piecewise-inextensible shape, with vertices and ridges whose size vanishes as $t\!\to \!0$.
Our next step is then to obtain a second, analogous criterion for the asymptotic description of diffused-stress structures.

As seen in the estimation of the energies, the diffuse-stress region
is characterized by a diverging aspect ratio: $W/L_t \to 0$ as $\tau \to 0$.
This feature is associated with a mechanical constraint on the configuration: a vanishing ratio between compressive and tensile stress components.  This is seen most clearly by considering the Airy potential $\chi(x,y)$ \cite{LL86}.
The scaling behavior leading to Eq.~(\ref{diffusedstressLtenergy}) suggests that in the limit $\tau \!\ll\!1$ Airy potentials of sheets with various thicknesses satisfy a scaling solution: $\chi(x,y) = \tau \tilde{\Delta} Y W^2  \ g (\bar{x},\frac{y}{W})$.  The coordinates $x$ and $y$ appear only in their rescaled forms, so $g$ depends only on the geometry of the system and boundary conditions. The prefactor is chosen to satisfy the scaling relations~(\ref{diffusedstressLtenergy}).
Because of the different scaling of $x$ and $y$ derivatives,
the asymptotic stresses scale as:
\begin{align}
\sigma_{xx} &\sim Y \tilde\Delta \tau & \sigma_{yy} &\sim Y \tilde\Delta \tau^2 & \sigma_{xy} \sim Y \tilde\Delta \tau^{3/2}.
\label{stressasym}
\end{align}
This asymptotic behavior is confirmed in Fig.~3,
and shows that as $\tau \! \to\! 0$ a single stress component $\sigma_{xx}$ becomes dominant. This is unlike
the asymptotic behavior of the strain, where Eqs.~(\ref{stressstrain},\ref{stressasym}) imply:
\begin{align}
u_{xx} &\sim \tilde\Delta \tau & u_{yy} &\sim -\nu \tilde\Delta \tau & u_{xy} &\sim \tilde\Delta \tau^{3/2} \ ,
\label{strainasym}
\end{align}
This leads us to conjecture that a diffuse-stress region is generally characterized by a vanishing stress ratio:
\begin{equation}
\sigma_{yy}/\sigma_{xx} \to 0 \ \ \text{as} \ \ \tau \to 0 \ ,
\label{constraintasym}
\end{equation}
regardless of the Poisson ratio. This mechanical property stands in contrast to the geometric, piecewise-inextensibility constraint that dominates focused-stress structures.
Condition~(\ref{constraintasym}) implies that the diffuse-stress region must terminate where the tensile stress component $\sigma_{xx}$ vanishes, as it does for sufficiently large $x$, in the single buckle region (whereas $\sigma_{yy}$ remains finite for any $\tau$ due to the confinement). This suggests that the focused-stress region appears precisely where condition (\ref{constraintasym}) can no longer be satisfied.

The vanishing stress ratio throughout the whole diffuse-stress region provides a basis for an asymptotic expansion of the shape in the limit $\tau \! \ll \! 1$. The simple geometry of our system enables us to demonstrate the basic principle of this expansion, since a natural candidate for the asymptotic diffuse-stress shape is obtained by a decomposition into two leading Fourier modes:
 \begin{equation}
 \zeta(\bar{x},y) \approx f_1(\bar{x})\cos(\pi y/W) + f_3(\bar{x})\cos(3\pi y/W)
 \label{asymshape}
  \end{equation}
for $0 < \bar x < \bar x^*$,
  where both functions $f_1(\bar{x}),f_3(\bar{x})$ remain finite and higher order modes (e.g. $f_n(\bar{x})$ with $n \!>\! 3$) vanish as $\tau \!\to\!0$. Intuitively, $f_3(\bar{x})$ is finite due to the imposed 3-buckle profile at $x\!=\!0$ whereas $f_1(\bar{x})$ must be finite in the approach to the single buckle shape. The ansatz~(\ref{asymshape}) is supported by our numerical data, Fig.~4.
  \begin{figure}
 \includegraphics{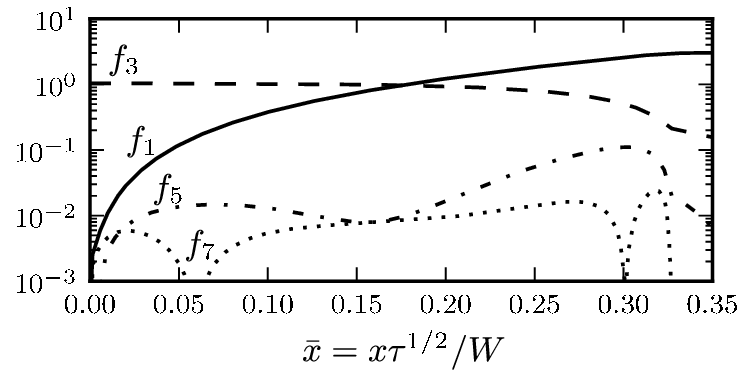}
 \caption{A linear-log plot of the first four odd Fourier modes of the cross-sections of the shapes for $t/W=0.0005$.
 Except near the focused structure at $\bar x^*\approx 0.32$, modes higher than 3 are negligible, indicating that most of the shape is smooth.}
 \end{figure}
  Obviously, the simple form~(\ref{asymshape}) does not describe the focused-stress region.
  The FvK equations~\cite{LL86} then provide a non-linear coupling between the shape $\zeta (\bar{x},y)$ and the Airy potential $\chi(\bar{x},y)$, and imply a similar asymptotic expansion for $\chi$ that exhibits precisely the vanishing stress ratio (\ref{constraintasym}) \cite{stressexpansion}. Equation~(\ref{asymshape}) results from the simple geometry of Fig.~1, but we expect that similarly smooth forms, composed of finite number of suitable basis functions, describe the shape and stress distribution of diffuse-stress domains in more complicated geometries.
  In our case, for $\tau \ll 1$, the FvK equations reduce to a set of coupled ordinary differential equations (ODE) for $f_1(\bar x)$ and $f_3(\bar x)$. We expect this reduction to ODE form to be characteristic of diffuse-stress domains. A full solution would require matching of the diffuse-stress shape (\ref{asymshape}) to the focused-stress structure at $\bar x^*$. The explicit form of the equations, as well as the effective matching conditions, will be discussed elsewhere \cite{inprep}.
 

We conclude by noting the possible relevance of our results for non-Euclidean elasticity,
which addresses the shape of sheets whose strainless state corresponds to some nonflat ``target" metric \cite{BenAmar08,Efrati09}.
Such sheets are naturally bent even in the absence of any external confining forces, and are
often assumed to be in a state that is close to ``isometric embedding", namely a strainless shape compatible with the target metric. This reasoning is reminiscent of focused-stress regions, wherein the sheet becomes strainless nearly everywhere except in ridges and vertices that become infinitely narrow zones whose area vanishes asymptotically.  However, there may be domains analogous to the diffuse-stress zones, where the strain becoming small everywhere in a region whose size is diverging.  Thus, several distinct expansions would need to be stitched together to describe the whole sheet.  It remains to be seen whether such a scenario emerges in non-Euclidean sheets.

In summary, we identified two classes of elastic building blocks that can be described as focused-stress and diffuse-stress, and showed how they coexist in elastic sheets under weak, smooth confinement.
These two classes are distinguished not only by their stress distributions, but also by their distinct characteristic energies and the different underlying asymptotic constraints: A geometric constraint (piecewise inextensibility) provides a theoretical framework for calculating shapes and energies of focused-stress domains, whereas a mechanical constraint (Eq.~(\ref{constraintasym})) is a basis for systematic analysis of diffuse-stress regions.
These observations were made by studying an elementary set-up that lead to coexistence between a single focused-stress domain and a single diffuse-stress region that channels stretching along a direction
dictated by the uniaxial confinement.
Further progress will be required
to develop
these concepts into a theoretical toolbox
that will allow efficient
analysis of thin sheets subject to general
types of forcing.

We thank A. Boudaoud and B. Audoly for useful discussions, and to B. Audoly for sharing with us an early draft
of \cite{AudolyBook}. We acknowledge support by NSF-MRSEC on Polymers at UMass (R.S.) and the Petroleum Research Fund of ACS (B.D.).  We thank the Aspen Center for Physics for its hospitality.


\begin{thebibliography}{22}
\bibitem{dconenote}
A d-cone characterized by a diverging length ($L_t \sim t^{-1/3}$) emerges in a pinched cylinder \cite{MahaPRL}. In contrast to our problem, pointwise pinching introduces a ``microscopic'' scale ($t$) to the geometry, and allows for $t \! \ll \! L_t \! \ll\! W$.
\bibitem{stressexpansion}
The corresponding expansion for the Airy potential is: $\chi(\bar x, \bar y\!\equiv\!\pi y/W) = \chi_0(\bar x)\bar y^4 + B(\bar x)\bar y^2 + D(\bar x) + \sum_{n=1}^3 \chi_{2n}(\bar x)\cos 2n\bar y$, where the $\chi_j$ can be expressed in terms of $f_1$ and $f_3$, and $B(\bar x)$ and $D(\bar x)$ are integration constants.
\end{thebibliography}
\end{document}